\documentclass[12pt]{article}

\begin{document}
\sloppy
\begin{flushright}{SIT-HEP/TM-4}
\end{flushright}
\vskip 1.5 truecm
\centerline{\large{\bf Affleck-Dine baryogenesis in the local domain}}
\vskip .75 truecm
\centerline{\bf Tomohiro Matsuda
\footnote{matsuda@sit.ac.jp}}
\vskip .4 truecm
\centerline {\it Laboratory of Physics, Saitama Institute of
 Technology,}
\centerline {\it Fusaiji, Okabe-machi, Saitama 369-0293, 
Japan}
\vskip 1. truecm
\makeatletter
\@addtoreset{equation}{section}
\def\theequation{\thesection.\arabic{equation}}
\makeatother
\vskip 1. truecm
\begin{abstract}
\hspace*{\parindent}
For Affleck-Dine baryogenesis to proceed, there must have been two
types of phase transitions.
One is the destabilized-stabilized phase transition of the flat
direction, which is in general induced by the Hubble parameter.
The other is the phase transition related to the A-term, which induces
the misalignment of the relative phase of the flat direction.
In the conventional Affleck-Dine baryogenesis they are supposed to 
start almost simultaneously.
Of course these phase transitions can take place separately, but the
latter must not be later than the former because the phase transition of
the A-term can not produce any baryon number when there is no
condensate of the relative charge.

In this paper we try to construct models where the original idea of 
Affleck-Dine baryogenesis is realized in a different way.
We show examples in which the {\it local} domain of the false vacuum with the
required condensate is formed {\it after} inflation and collapses in a
safe way so that the domain wall problem is avoided.
We also show examples where the phase transition of the A-term starts
before the decay of the condensate.
As in the conventional Affleck-Dine mechanism, the phase transition of
the A-term produces baryon number in the {\it local} domain of the
 condensate. 
We construct scenarios where our mechanism produces sufficient baryon 
asymmetry of the Universe.
\end{abstract}

\newpage
\section{Introduction}
\hspace*{\parindent}
The production of net baryon asymmetry requires baryon number violating
interactions, C and CP violation and a departure from thermal 
equilibrium\cite{sakharov}.
In the original and simplest model of baryogenesis\cite{original}
a GUT gauge or Higgs boson decays out of equilibrium producing
a net baryon asymmetry.
Another mechanism of generating the cosmological baryon asymmetry
in supersymmetric theories is proposed by Affleck and 
Dine\cite{AD} who utilized the decay of scalar condensate along the
flat direction. 
This mechanism is a natural product of supersymmetry, which contains
many flat directions that break $U(1)_{B}$.
The scalar potential along this direction vanishes identically when 
supersymmetry breaking is not induced.
Supersymmetry breaking lifts this degeneracy,
\begin{equation}
V\simeq m^{2}_{soft}|\phi|^{2}
\end{equation}
where $m^{2}_{soft}$ is the supersymmetry-breaking scale and
$\phi$ is the direction in the field space corresponding to
the flat direction.
For large initial values of $\phi$, a large baryon number asymmetry
may be generated if the condensate of the field breaks $U(1)_{B}$.
The mechanism also requires the presence of baryon number violating 
operators that may appear through higher dimensional A-terms.
The decay of these condensates through such an operator can lead to a 
net baryon asymmetry.
In the most naive consideration the baryon asymmetry is computed by 
tracking the evolution of the sfermion condensate.
Considering a toy model with the potential
\begin{equation}
V(\phi,\phi^{\dagger})=m^{2}_{soft}|\phi|^{2}+\frac{1}{4}[\lambda
\phi^{4}+h.c.],
\end{equation}
the equation of motion becomes
\begin{equation}
\label{toy}
\ddot{\phi} + 3H\dot{\phi} = - m_{soft}^2 \phi 
+ \lambda (\phi^{\dagger})^{3}.
\end{equation}
The baryon (or lepton) number density is given by:
\begin{equation}
n_{B}=q_{B}\left(\dot{\phi}^{\dagger}\phi-\phi^{\dagger}\dot{\phi}\right),
\end{equation}
where $q_{B}$ is the baryon (or lepton) charge carried by the field.
Now one can write down the equation for the baryon number density
\begin{equation}
\label{nbeq}
\dot{n_{B}}+3Hn_{B}=2q_{B}Im\left[\lambda (\phi^{\dagger})^{4} 
\right].
\end{equation}
Integrating this equation, one can obtain the baryon (or lepton)
number produced by the Affleck-Dine oscillation.
For large initial amplitude, the produced baryon number is estimated as
$n_{B} \simeq \frac{4q_{B}|\lambda|}{9H} |\phi_{ini}|^4 \delta_{eff}$,
where $\delta_{eff}$ is the effective CP violation phase of the initial
condensate.
This crude estimation suggests that by generating some angular motion 
one can generate a net baryon density.

In conventional mechanism of Affleck-Dine baryogenesis, one should
assume large $H>m_{3/2}$ before the time of  Affleck-Dine baryogenesis to
destabilize the flat directions and obtain the large initial amplitude
of baryon-charged directions.

Although it seems plausible that Affleck-Dine baryogenesis generates
baryon asymmetry of the Universe, there are some difficulties in the 
naive scenario.
The formation of Q-ball\cite{kusenko} is perhaps the most serious obstacles
that puts a serious constraint on the baryon number density at the
time of Q-ball formation.
Q-balls are formed due to the spatial instability of the Affleck-Dine
field, and have been shown by numerical calculations that they absorb
almost all the baryonic charges in the Universe when they
form\cite{Kasuya, Enq}.
This means that the baryon asymmetry of the Universe in the later period
must be provided by decaying Q-balls.
In general, the stability of Q-ball is determined by their charge,
which is inevitably fixed by Affleck-Dine mechanism itself.
The reason is that the formation of Q-ball occurs almost immediately,
which makes it hard to expect any additional diluting mechanism before
Q-ball formation.
The point is that in general Affleck-Dine baryogenesis the initial
baryon number density becomes so huge that the produced Q-balls are stable.
The stable Q-balls that produce the present baryon asymmetry of the
Universe by their decay are dangerous, because such Q-balls can also produce
dangerous relics at the same time when they decay to produce the baryons.
The decay temperature of the associated Q-balls becomes in general 
much lower than the 
freeze-out temperature of the dangerous lightest supersymmetric
particle, which causes serious constraint.

Another obstacle is the problem of early oscillation caused by the
thermalization\cite{dineT}.
When the fields that couple to the Affleck-Dine field is thermalized,
they induce the thermal mass to the Affleck-Dine field.
The early oscillation starts when the thermal mass term exceeds the
destabilizing mass.
The serious constraint appears because the destabilizing mass, which
is about the same order of the Hubble parameter, is in general 
much smaller than the temperature of the plasma.

Now let us explain what we have in mind.
In general Affleck-Dine baryogenesis the following conditions are
required to produce net baryon asymmetry.
\begin{itemize}
\item There exists the field condensate $\phi_{AD}\ne 0$ charged with the
      baryon number. The condensate must decay during or after the
      baryon number production. 
\item The A-term that determines the initial phase of the condensate
      must be 
      different from the one in the true vacuum. There must be the phase
      transition of the A-term that rotates the phase of the
      condensate. The rotation produces net baryon number.
\end{itemize}
In conventional models of Affleck-Dine baryogenesis it is assumed that
the phase transition of the collapsing condensate and the phase
transition of the A-term take place almost simultaneously during the
inflaton-dominated era of the Universe.
Both of them were supposed to be induced by the supersymmetry breaking
of the decreasing Hubble parameter.
In the scenario of the conventional Affleck-Dine baryogenesis, the
A-term induced by non-zero $H$ displaces the CP-violating phase.
As $H$ becomes smaller, other sources for the A-term dominate and
the shift of the phase turns into the driving
force in the angular direction, which produces net baryon number.

However, in some cases, these phase transitions do not start
simultaneously.
The most serious example is the early oscillation induced by the thermal
mass of the flat direction, which starts before the phase transition of
the A-term\cite{dineT}.
In such cases, the required baryon number is not produced.
On the other hand, one can remedy this situation if the phase transition
of the A-term starts earlier than the collapse of the
condensate\cite{dineT}.

Here we try to construct models where the required phase transitions are
completely separated from the inflation.
When one discusses the cosmological domain wall, one must take into
account the large (but local) domains of the quasi-degenerated vacua.
In generic situations the radius of the domain immediately becomes as
large as the Hubble radius.
If the required condensate exists in one of these domains, the first
requirement $\phi_{AD}\ne 0$ is satisfied locally in the domain.
Then our task is to construct a model where the A-term changes its
phase before the collapse of the quasi-degenerated domain of the
condensate.\footnote{The criteria for the cosmological domain wall is
discussed in ref.\cite{KZO}. If they ever existed, they must have disappeared
at $t<t_c \sim (G\sigma)^{-1}$\cite{vilenkin}.}

The most significant difference from the conventional Affleck-Dine
baryogenesis is that the production of the baryon number starts at
much later period of the evolution of the Universe.

In this paper we construct a concrete model in which the required
condensate is realized in the domain of the quasi-degenerated vacua.
Our model starts with the situation where the Universe is occupied by the
domains of the false and the true vacua.\footnote{
Of course one may consider the case where
the destabilization of the field (\ref{omit}) appears inside 
cosmological defects such as walls and strings.  
Then there will be the volume factor suppression.
The volume suppression is $\epsilon_{vol}\sim \Delta_{AD}/H^{-1}$ for
domain walls and $\epsilon_{vol}\sim (\Delta_{AD}/H^{-1})^{2}$ for
strings, if one assumes only one configuration per unit horizon.
For fat walls and strings, a viable baryon number
production is expected in the allowed parameter
regions\cite{matsuda_tobe}.} 
We also construct models in which the phase transition of the A-term is
realized at low energy scale.

Let us consider a toy model with a potential in the domain of the 
false vacuum,
\begin{equation}
\label{omit}
V(\phi, \phi^{\dagger})=-m^{'2}|\phi_{AD}|^{2}
+\left[\frac{a_{o}}{4M_{p}}\phi^{4}_{AD}+h.c.\right]
\end{equation}
where the vacuum expectation value of the Affleck-Dine field $\phi_{AD}$ 
is non-zero.
$a_{o}$ represents the A-term in the true vacuum.\footnote{
Eq.(\ref{omit}) represents the potential in the false vacuum domain of
the condensate {\it after} the A-term phase transition.
Please note that we are considering two types of false vacuum 
configurations that appear independently.
}
We must also include the temporal A-term,
\begin{eqnarray}
\label{tempo}
V_{A}(\phi, \phi^{\dagger})=\left[\frac{a'}{4M_{p}}\phi^{4}+h.c.\right]
\end{eqnarray}
which disappears after $t>t_{A}$.
Here $t_A$ denotes the time when the phase transition of the A-term
takes place, and $a'$ is induced by the false vacuum configuration
during $t<t_A$.
$a'$ is required to be larger than $a_{o}$.
Integrating the equation\footnote{
See eq.(\ref{nbeq}) in the review part.}, 
one can find the produced
baryon number density in the domain of the false vacuum,
\begin{equation}
\label{18}
n_{B}\sim \frac{4q_{B}}{9H}\left| \frac{a_0}{M_p} 
(\phi_{AD}^{ini})^4 \right|\delta_{eff}.
\end{equation}
Unlike the conventional Affleck-Dine baryogenesis, here the Hubble parameter
$H$ is the parameter that should be determined by the time of the
A-term phase transition at $t=t_{A}$.

In Section 2 we show example in which the condensate
is realized inside the domain of the quasi-degenerated vacua.

In section 3 we consider  
examples in which the misalignment of the A-term appears at the phase
transition of the A-term.

\section{Condensate in the false vacuum domain}
\hspace*{\parindent}
In general models of Affleck-Dine baryogenesis, the destabilization
of the flat direction is expected to be induced by the Hubble parameter, 
which is expected to be larger than $m_{3/2}$ at an earlier stage of 
the Universe.

In this section we consider examples in which the Affleck-Dine condensate 
is realized inside the domain of the false vacuum.
The most significant difference from the conventional scenario
is that the condensate is realized in the local domain and it can
remain at much lower temperature {\it after} reheating.

We consider an additional sector that is described by the
superpotential
\begin{equation}
\label{simplest}
W_{M}=\frac{1}{3}\lambda S^{3}-\kappa_{s}S\Lambda^{2}+
\kappa_{Q}S \overline{Q}^{i}Q^{i}.
\end{equation}
The quark-like superfield $Q^i$ is not a conventional ingredient of the
standard model, but charged with the standard model gauge group and
$U(1)_{B}$. 
$Q^i$ can be replaced by the lepton-like superfield $L^i$ if one considers
Affleck-Dine leptogenesis.
The minima of this potential are at
\begin{eqnarray}
\label{first}
<S>&=&\pm \sqrt{\frac{\kappa_{s}}{\lambda}}\Lambda\nonumber\\
<\overline{Q}^{i}Q^{i}>&=&0
\end{eqnarray}
and 
\begin{eqnarray}
\label{second}
<S>&=&0\nonumber\\
<\overline{Q}^{i}Q^{i}>&\sim& \Lambda^{2}.
\end{eqnarray}
We set the scale $\Lambda$ at the intermediate scale
$M_{EW}\ll\Lambda\ll M_{p}$, and assume that
$\kappa_{s}$, $\kappa_{Q}$ and $\lambda$ are constants of O(1).

In this case, the Affleck-Dine field is {\it not} intended to be the flat 
direction of the model. 
The Affleck-Dine field is the baryonic direction that develops non-zero 
vacuum expectation value in the false vacuum (\ref{second}).
Here $N_{f}\ge N_{c}$ heavy quarks are required in our messenger sector,
which decouple in the true vacuum while they condensate in the false vacuum. 
We can define baryonic combinations  
$B^{i_{1}...i_{N_{c}}}=Q^{i_{1}}...Q^{i_{N_{c}}}$
as in the usual models of supersymmetric QCD, which is schematically 
denoted by $\phi_{AD}$. 
Obviously, this direction is not the flat direction of the model.
One can also add heavy leptonic fields $L^i$ in this sector.

Here we make a brief comment on the stability of the false vacuum and
related domain walls.
First, the degeneracy between (\ref{first}) and (\ref{second}) is obviously
broken by the soft mass terms.
In the presence of the soft mass terms, the difference between these vacua
becomes $\epsilon \sim m_{soft}^2 \Lambda^2$ if there is no miraculous
cancellation, which is enough to explain
the safe decay.
We should also consider the degeneracy between 
$S=\pm \sqrt{\frac{\kappa_{s}}{\lambda}}\Lambda$.
In this case, one may gauge the relevant discrete symmetry, or expect
destabilization by the effective terms induced by the gravitational
interactions.

\section{A-term phase transition}
\hspace*{\parindent}
In this section we construct several examples of the
A-term phase transition that induces the driving force in the angular
direction of $\phi_{AD}$.

\subsection{Thermal phase transition}
\hspace*{\parindent}
We shall first consider an example where the thermal effect triggers 
the phase transition at the critical temperature $T_c$.

The phase transition can be realized by imposing extra gauge
symmetries such as $U(1)_{B-L}$ or other string-motivated $U(1)$ 
symmetries.\footnote{
Affleck-Dine baryogenesis with gauged $U(1)_{B-L}$ is discussed in
ref.\cite{b-l} from a different point of view.}
These extended gauge symmetries are sometimes utilized to certificate
the proton decay in many kinds of grand unification models.
We can consider the case where the higher dimensional terms that
break baryon (or lepton) number are suppressed by enhanced gauge
symmetries\cite{pati,z_symmetry}, which is sometimes discussed in the light
of string-derived GUT models.
These symmetries are used to suppress naturally the unsafe
$d=4$ as well as the color-triplet mediated or gravity induced
$d=5$ proton-decay operators, which generically arise in unification
models of supersymmetry.
For example, in the models that contain $B-L$ as a symmetry, 
the dangerous $d=4$ operators permitted by the standard model gauge 
symmetry are forbidden.
However, such operators can in general appear through non-renormalizable
operators if there exists a spontaneous breaking of the symmetry
in the presence of non-zero vacuum expectation values of
the fields that carry $B-L$.
To eliminate the dangerous $d=5$ operators, one may add other
additional gauge symmetries or adjust the charge assignment of the
$B-L$ breaking field.
Of course one may use discrete gauge symmetries
alternatively\cite{discrete}. 

Here we consider an additional superfield $Z$ charged with
$U(1)_{B-L}$. 
We assume that $Z$ is a standard model singlet with the flat potential
destabilized by the soft mass.
We consider a nonrenormalizable operator that contains $Z$,
\begin{equation}
\label{nonre0}
W_{A}\sim \frac{1}{M_{p}^{n-3}}\left(\frac{Z}{M_{p}}\right)\phi_{AD}^{n}
\end{equation}
where $\phi_{AD}$ denotes the AD direction that carries non-zero
$U(1)_{B-L}$ charge.
We have set that the cut-off scale is simply given by the Planck scale.
The relevant A-term takes the following form
\begin{equation}
V_{A}\sim a m_{3/2} \frac{1}{M_{p}^{n-3}}
\left(\frac{Z}{M_p}\right)\phi^n_{AD}+h.c.
\end{equation}
where $a$ is the constant of $O(1)$.

$Z$ can be thermalized to obtain the thermal mass $\sim T^2$,
which stabilizes the potential.
Let us assume that the above superpotential dominates the A-term in the
{\it true} vacuum.
On the other hand, in the {\it false} vacuum where $Z$ vanishes, 
other terms can naturally contribute to determine the phase of the
condensate  
so that the expected value of the phase becomes
different from the one in the true vacuum.\footnote{
Without loss of generality, we can assume that there are other sources of
A-term that determine the phase of the condensate when $Z$ vanishes.
For example, there is the A-term of the form 
$V_A\sim \frac{F_Z}{M_p^{n-2}}\phi_{AD}^n+h.c.$ where $F_Z\sim
\rho_I^{1/2}$ during thermal inflation.
} 
In this respect, the gap of the phase appears between the above two vacua.
When the false vacuum of the A-term eventually collapses, the
misalignment of the 
phase of $\phi_{AD}$ produces net baryon number in the local
domain of the condensate $\phi_{AD}\ne 0$.
In this model, the phase transition from $<Z>=0$ to $<Z>\simeq\Lambda$ 
takes place at the temperature about $T_{c}\sim
m_{3/2}$ since the direction is expected to be flat around the
origin.\footnote{
The nearly flat direction of this form may induce thermal
inflation\cite{thermal}.
The evolution of the domain wall during thermal inflation is already
discussed in ref.\cite{matsuda_thermal_wall}.}
In this case, there should be a short period of thermal inflation,
and the baryon number production occurs after thermal inflation.
We can estimate the baryon to entropy ratio\cite{AD},
\begin{equation}
\frac{n_{B}}{s} \sim\frac{T_{R_Z}}{H_o}
\frac{|a m_{3/2} Z \phi_{AD}^n|}{M_p^{n-2}\rho_I}
\end{equation}
where $\frac{n_b}{n_{\phi_{AD}}}$ is the ratio of the baryon to Affleck-Dine
field number densities, which can be assumed to be $\sim O(1)$.
$H_o$ denotes the Hubble parameter at the time when the oscillation of
the AD field starts, and
$T_{R_Z}$ is the reheating temperature after thermal inflation.
Assuming that the energy density during thermal inflation $\rho_{I}$ is
$\rho_I \sim (m_{3/2} Z)^2$, we can obtain the crude
estimation for $n=4$, 
\begin{equation}
\frac{n_{B}}{s} \sim 10^{-10}\left( \frac{T_{R_Z}}{1 GeV} \right)
\left(\frac{10^2 GeV}{m_{3/2}}\right)^{2}
\left(\frac{10^{12}GeV}{<Z>}\right)^{2} 
\left(\frac{\phi_{AD}}{10^{9} GeV}\right)^4,
\end{equation}
where $<Z>$ is the vacuum expectation value of $Z$ in the true vacuum.
Although in general $H_o$ is expected to be less than $H_I=\rho_Z^{1/2}/M_p$, 
here we have assumed $H_o \simeq H_I$ which
corresponds to the case where the AD oscillation starts immediately
after thermal inflation.

In this case, the baryon number is actually generated.

Let us discuss the string-derived symmetries for an another example.
Without any additional symmetry other than the standard model gauge
symmetry, a supersymmetric theory in general permits dimension 4 and 
dimension 5 operators that violate the baryon and lepton numbers.
The operators in problem are written by using standard notations,
\begin{eqnarray}
W&=&[\eta_{1}\overline{U}\overline{D}\overline{D}+\eta_{2}QL\overline{D}
+\eta_{3}LL\overline{E}]\nonumber\\
&=&\frac{1}{M}[\lambda_{1}QQQL + \lambda_{2}\overline{U}\overline{U}
\overline{D}\overline{E}+\lambda_{3}LLH_{2}H_{2}].
\end{eqnarray}
Experimental limits on proton lifetime impose the constraint on the 
couplings as:
$\eta_{1}\eta_{2}\le 10^{-24}$ and $\lambda_{1,2}/M \le
10^{-25}(GeV)^{-1}$.
One may also impose the requirements that the symmetry naturally ensures
the proton stability and simultaneously permits the light neutrino
masses.
The solution given in ref.\cite{pati,z_symmetry} utilizes the gauged $B-L$ and
other string-motivated symmetries, such as
\begin{equation}
\label{sthid}
\lambda_{1} \sim 
\left(\frac{\tilde{\overline{N}_{R}}}{M}\right) \left( \frac{T_{i}
\overline{T_{j}}}{M^{2}}\right)^{2}\sim 10^{-2.5}\left( \frac{\Lambda_{c}}
{M}\right)^{4}
\end{equation}
where $T_{i}\overline{T_{j}}$ is the condensate in the hidden sector and
their magnitude is denoted by $\Lambda_{c}^{2}$.\footnote{
See ref.\cite{pati} for notations.}
The experimental bound on $\eta_{1,2}$ implies $(\Lambda_{c}/M)^{4}
\le 10^{-9.5}$.
On the other hand, $\Lambda_{c}$ cannot be less than $TeV$, since then
the heavy particles of the relevant gauge symmetry would be accessible
to the experiments. 

Here we consider another operator  
\begin{equation}
W_{A}=\lambda_{A}\frac{T_i\overline{T}_j}{M_p^2} \frac{\phi_{AD}^4}{M_p}.
\end{equation}
In this model, naive thermal phase transition in the hidden sector 
($T_{i}\overline{T_{j}}=0$ to $T_{i}\overline{T_{j}}=\Lambda_c^2$)
is expected to induce the misalignment of the phase.
In this case, the temperature of the baryon number production is
determined by the scale of the condensate of the hidden sector matter
field $T_i$. 
The resultant baryon number is\footnote{
See eq.(\ref{18}) in the introduction.}
\begin{equation}
n_{B} \sim \frac{1}{H} \left[m_{3/2}\frac{|T_i\overline{T}_j
\lambda_A \phi_{AD}^4|}{M_p^3} \right]
\end{equation}
where we have omitted numerical factors and parameters of $O(1)$.
We estimate the  baryon to entropy ratio at the critical temperature
$T_c \sim \Lambda_c$,
\begin{equation}
\frac{n_{B}}{s} \sim 10^{-10} 
\left(\frac{\phi_{AD}}{10^{10}GeV}\right)^4
\left(\frac{10^{6} GeV}{\Lambda_c}\right)^3
\left(\frac{m_{3/2}}{10^2 GeV}\right).
\end{equation}

\subsection{Less trivial example}
\hspace*{\parindent}
Here we consider different types of phase transitions.
In both cases, misalignment of the phase is induced by the changes in
the effective A-term.
The difference is that in the above examples the changes are induced by
thermal phase transition while in the following examples they are
consequences of the decay of the false vacuum domains of the A-term.
In general, there are two types of such phase transitions.
In one case, an alternative A-term becomes larger than the ordinary one, 
while in the other case the misalignment appears because the phase of
the conventional A-term takes different values in the
quasi-degenerated domains. 

\underline{Larger A-term appears in the false vacuum}

The authors of ref.\cite{ellis} argued that additional A-terms 
possibly appear by thermal effect and it will be proportional to 
the temperature $T$.
In ref.\cite{dineT}, however, it is discussed that the A-term will not
appear in generic situation.

Here we consider an alternative of A-term generation in
ref.\cite{ellis}.
We consider the case where the condensate of the trigger field is not
due to the thermal effect, but is realized by the false vacuum field
configuration.
We consider terms in the potential of the form:
\begin{equation}
W=\lambda\frac{\phi_{AD}^{4}}{M}+h\phi_{AD}\xi\xi,
\end{equation}
where the field $\xi$ represents extra matter field that
couples to the Affleck-Dine field with a Yukawa coupling.
Higher dimensional couplings may be included in the coupling constant
$\lambda$ and $h$.
Then the potential includes terms such as:
\begin{equation}
V_{A}=h\lambda\frac{\phi_{AD}^{3}}{M}\xi^{*}\xi^{*}.
\end{equation}
If large vacuum expectation value of $<\xi^*\xi^*>$ is formed in the
false vacuum, quite large A-term will effectively appear.
As we have discussed in the previous section, it is possible to 
construct such configurations in which the scalar component of
the field $\xi$ develops large expectation value in the false vacuum.
When the false vacuum bubble shrinks or collapses 
due to the energy difference, misalignment will appear.\footnote{
Here the false vacuum of the A-term does not cover the whole Universe,
but it appears 
as a false vacuum domain of the $H^{-1}$ size.
Baryogenesis is expected to take place in the regions where both
the field condensate and the additional A-term exist.
}
We can estimate the produced baryon number,
\begin{equation}
n_{B} \sim \frac{1}{H}\left[
\lambda \frac{m_{3/2}}{M_{p}}\phi_{AD}^4
\right]
\end{equation}
where numerical factors and parameters of $O(1)$ are ignored.
The resultant baryon to entropy ratio is
\begin{equation}
\frac{n_{B}}{s}\sim 10^{-10} 
\left(\frac{\lambda}{10^{-11}}\right)
\left(\frac{m_{3/2}}{10^{2}GeV}\right)
\left(\frac{\phi_{AD}}{10^6 GeV}\right)^4
\left(\frac{10^5 GeV}{T_{A}}\right)^5
\end{equation}
where $T_{A}$ denotes the temperature when the condensate of $\xi$ decays.

For an alternative example, here we consider a model with the
superpotential 
\begin{equation}
W=S(\overline{X}X-\Lambda_{X}^{2})+S^{3}.
\end{equation}
Here we assume that $S$ is charged with the discrete gauged R-symmetry.
The two degenerated minima of this potential are
\begin{eqnarray}
\label{falseBL}
<S>&\simeq& \pm \Lambda_{X} \nonumber\\
<\overline{X}X>&=&0
\end{eqnarray}
and
\begin{eqnarray}
\label{trueBL}
<S>&=&0\nonumber\\
<\overline{X}X>&\simeq&\Lambda_{X}^{2}.
\end{eqnarray}

The nonrenormalizable operators in the superpotential can be written as
\begin{equation}
\label{nonre}
W_{A}\simeq \frac{1}{M_{p}^{n-3}}
\left(\frac{S}{M_{p}}\right)\phi_{AD}^{n}
\end{equation}
where $\phi_{AD}$ denotes the AD direction.
We have set that the cut-off scale is simply given by the Planck scale.

When the false vacuum of $S=0$ eventually collapses, the
misalignment of the  
AD phase appears producing net baryon number in the local
domain of the AD condensate.
In general the temperature of the baryogenesis is determined by the
scale of $\Lambda_{X}$, from the requirement that the cosmological
domain walls must decay before they dominate the Universe.\footnote{
The constraint is discussed in appendix A.}
For $n=6$, the produced baryon number is 
\begin{equation}
n_{B} \sim \frac{1}{H}\left[
\frac{m_{3/2}}{M_{p}^3}\frac{\Lambda_{X}}{M_p}\phi_{AD}^6
\right]
\end{equation}
where numerical factors and parameters of $O(1)$ are ignored.
The resultant baryon to entropy ratio is
\begin{equation}
\frac{n_{B}}{s}\sim 10^{-10} 
\left(\frac{\Lambda_{X}}{10^{9}GeV}\right)
\left(\frac{m_{3/2}}{10^{2}GeV}\right)
\left(\frac{\phi_{AD}}{10^9 GeV}\right)^6
\left(\frac{10^4 GeV}{T_{A}}\right)^5
\end{equation}
where $T_{A}$ denotes the temperature when the false vacuum domain
of the A-term collapses.

\underline{Misalignment of the A-term}

We can also consider another example in which the domain wall 
appears simply as the consequence of the spontaneously broken $Z_{N}$ 
symmetry of the coefficient in the A-term, which may appear when the
A-term is determined 
by the field with the effective $Z_N$ symmetry.
Then the misalignment of the A-term appears as the consequence of the
discrete phases of a field, 
for example $<X> \simeq \Lambda_{X}e^{\frac{2\pi i k}{N}}$.
The appearance of such misalignment is very natural since each field
appearing in the A-term may have their effective discrete symmetries that
is spontaneously broken by their vacuum expectation values.

In this case, there is a naive concern that the produced baryon
in each $k$-domain may cancel when the contributions are summed over the
whole Universe.
To see what happens in our model, let us consider the relevant part of
the potential of the form
\begin{equation}
V_{A} \sim \frac{\lambda a_{k}}{M_{p}^{n-3}}\phi^{n}_{AD}+h.c.
\end{equation}
where $a_k$ is assumed to be determined by the vacuum expectation value
of the field.
Here we consider the case where the misalignment occurs because 
the discrete domains  
of $a_{k}=\Lambda_{a}e^{\frac{2\pi i}{N}k}$ are formed before baryon
number production.
In addition to the domains of $k$-vacuum of the A-term, there are
discrete $n$ vacua that appear because 
of the effective $Z_{n}$ symmetry of the AD field.
As the result, the initial phase in each domain takes the form
\begin{equation}
\theta_{\phi_{AD}}^{ini} \simeq \frac{1}{n}\left[-arg(\lambda)-arg(a_k)
+ 2 \pi j \right],
\end{equation}
where $j=1,2,...,n$ and $k=1,2,...,N$.
In the conventional Affleck-Dine baryogenesis, one of the $(j,k)$ vacuum
dominates the whole Universe because of the inflationary period, which
determines the initial condition of Affleck-Dine baryogenesis.
On the other hand, in our case, the resultant baryon number may contain
the following summation factor.
Let us denote the A-term in the final state by $a_{k_0}$.
If there is no biasing, it is given by
\begin{eqnarray}
\epsilon_{sum} &=& \sum_{All\, Domain} 
\sin\left(arg(\lambda)+arg(a_{k_0})+n \, arg(\phi_{AD})\right)
\nonumber\\
&=&  \sum_{k=1}^{N} 
\sin\left(arg(\lambda)+arg(a_{k_0})+\left[-arg(\lambda)-arg(a_k)
+ 2 \pi j \right]\right)\nonumber\\
&=&  \sum_{k=1}^{N} 
\sin\left(arg(a_{k_0})-arg(a_k)\right)\nonumber\\
&=& 0
\end{eqnarray}
In the above example, cancellation occurs if there is no biasing at
the domain formation.\footnote{The mechanism for biasing the domain
formation is  
discussed by many authors.
See ref.\cite{bias} for example.
Estimation of the biasing factor requires computational simulations,
which is not suitable for our motivation for this paper.
In models for baryogenesis, the biasing factor is sometimes related to the CP
violating phase of O(0.01).}
On the other hand, however, complete cancellation is rather miraculous
because the biasing is expected in more realistic situations.
We expect the biasing factor of $\epsilon_{sum}\sim 0.1 - 0.01$ because
the vacua as well as the potential is not strictly symmetric in our
model.

Note that the cancellation cannot occur when the initial AD phase
is determined by other A-terms.

\section{Conclusions and Discussions}
\hspace*{\parindent}
In this paper we have tried to separate the mechanism of Affleck-Dine
baryogenesis from the inflation in order to avoid several obstacles in the
original model. 
We have shown several examples for realizing 
Affleck-Dine baryogenesis in the alternative way.
Our mechanism is quite efficient in producing baryons and
we believe that application of our mechanism may solve the
problems of the conventional models of baryogenesis.
The idea of Affleck-Dine baryogenesis proceeding with different
efficiency in local domains was suggested in ref.\cite{DS} from
different context.

\section{Acknowledgment}

We wish to thank K.Shima for encouragement, and our colleagues in
Tokyo University for their kind hospitality.

\appendix
\section{Constraint on the cosmological domain walls}
\hspace*{\parindent}
It is well known  that whenever the Universe undergoes a phase
transition associated with the spontaneous symmetry breaking,
domain walls will inevitably form.
In most cases, the domain walls are dangerous for the standard evolution
of the universe\cite{KZO}.
However, if some criteria are satisfied, unstable domain walls that
disappear at $t<t_c \sim (G\sigma)^{-1}$ can exist.
First, we review how to estimate the constraint
to safely remove the cosmological domain walls\cite{vilenkin}.
When the discrete symmetry is broken  by 
gravitational interactions, the symmetry is an approximate
discrete symmetry.
The degeneracy is broken and the energy difference 
$\epsilon\ne0$ appears.
Regions of the higher density vacuum tend to collapse,
the corresponding force per unit area of the wall 
is $\sim \epsilon$.
The energy difference $\epsilon$ becomes dynamically 
important when this force becomes comparable to the force of 
the tension $f\sim \sigma/R_{w}$,
where $\sigma$ is the surface energy density of the wall
and $R_{w}$ denotes the typical scale for the wall distance.
For walls to disappear, this has to happen before they become harmful.
On the other hand, the domain wall network is not a static 
system.
In general, the initial shape of the walls right after the
phase transition is determined by the random variation
of the scalar VEV.
One expects the walls to be very irregular, random
surfaces with a typical curvature radius, which
is determined by the correlation length of the
scalar field.
To characterize the system of domain walls,
one can use a simulation\cite{simulation}.
The system will be dominated by one large (infinite size)
wall network  and some finite closed walls (cells) when
they form.
The isolated closed walls smaller than the horizon
will shrink and disappear soon after the phase transition.
Since the walls smaller than the horizon size 
will efficiently disappear so that only walls
at the horizon size will remain, 
their typical curvature scale will be  the horizon 
size, $R\sim t\sim M_{p}/g_{*}^{\frac{1}{2}}T^{2}$.
Since the energy density of the wall $\rho_{w}$ is about
\begin{equation}
\rho_{w}\sim \frac{\sigma}{R},
\end{equation}
and the radiation energy density $\rho_{r}$ is 
\begin{equation}
\rho_{r}\sim g_{*}T^{4},
\end{equation}
one sees that the wall dominates the evolution 
below a temperature $T_{w}$
\begin{equation}
T_{w}\sim \left(\frac{\sigma}{g_{*}^{1/2}M_{p}}
\right)^{\frac{1}{2}}.
\end{equation}
To prevent the wall domination, one requires the
pressure to have become dominant before this epoch,
\begin{equation}
\label{criterion}
\epsilon>\frac{\sigma}{R_{wd}}\sim
\frac{\sigma^{2}}{M^{2}_{p}},
\end{equation}
which is consistent with the criterion in ref.\cite{KZO,vilenkin}.
Here $R_{wd}$ denotes the horizon size at the wall domination.
A pressure of this magnitude would be produced by
higher dimensional operators, which explicitly break
 the effective discrete symmetry\cite{matsuda_thermal_wall,matsuda}.

The criterion (\ref{criterion})
seems appropriate, if the scale of the wall is higher 
than $(10^{5}GeV)^{3}$.
For the walls below this scale ($\sigma\le(10^{5}GeV)^{3}$),
 there should be  further constraints coming from primordial 
nucleosynthesis.
Since the time associated with the collapsing temperature
 $T_{w}$
is $t_{w}\sim M_{p}^{2}/g_{*}^{\frac{1}{2}}\sigma
\sim 10^{8}\left(\frac{(10^{2}GeV)^{3}}{\sigma}\right)$sec,
the walls $\sigma\le(10^{5}GeV)^{3}$ will decay after 
nucleosynthesis\cite{Abel}.
In this case, one must consider stronger constraint.
If the walls are not hidden and are supposed to decay into standard
model particles, the entropy produced when walls collapse
will violate the phenomenological bounds for nucleosynthesis.
On the other hand, this simple bound
from the nucleosynthesis is not effective 
for the walls that cannot decay into standard model 
particles.
The  walls such as soft domain walls\cite{soft_wall},
the succeeding story should strongly depend on the 
details of the hidden components and their interactions.
These walls can decay late to contribute to the large scale
structure formation.

Of course, the condition for the cosmological domain wall not to dominate
the Universe (\ref{criterion}) 
should also be changed if the wall velocity is lower than
the speed of the light and then the Universe contains walls more than one. 
This implies that the condition to evade the wall domination becomes 
$\epsilon>(\sigma^{2}/M_{p}^{2})\times x$,
where the constant $x$ is determined by $R_{w}$ as 
$x\simeq M_{p}/(R_{w}T^{2})$.
For the walls with lower velocity, the bound for $\epsilon$ 
is inevitably raised since such walls will dominate earlier.

\section{Comment on the early oscillation}
\hspace*{\parindent}
In conventional scenarios of Affleck-Dine baryogenesis, the oscillation of 
condensates along flat directions that carry non-zero baryonic charge
and attained large vacuum expectation values at the end of the
inflationary epoch is important.
Although these directions are flat, they may couple to other fields
that acquire masses induced by the vacuum expectation values of the
flat directions.
If the masses of other fields are sufficiently small, the plasma of
inflaton and their decay products can act on the flat directions,
which may induce thermal masses.
In such cases, the flat directions acquire large masses and start their
oscillations earlier than usually estimated.
There will be other way terminating the oscillations due to evaporation
of the condensate.
These thermal effects should alter the estimates of the conventional
Affleck-Dine baryogenesis as is discussed in ref.\cite{dineT,ellis}.

In our paper we have discussed the phase transition of the A-term
at low energy scales.
However, the scales of the phase transitions can be raised
to much higher energy scales.
If the typical scales of such A-term phase transitions are raised so
that they can take place before the early oscillation,
they can produce baryon number before the problematic early oscillation
avoiding the problem of the original Affleck-Dine
baryogenesis\cite{matsuda_tobe}.


\begin{thebibliography}{1}
\bibitem{sakharov}
A.D.Sakharov, JETP Lett.5(1967)24
\bibitem{original}
S.Weinberg, Phys.Rev.Lett.42(1979)850\\
D.Toussant, S.B.Treiman, F.Wilczek and A.Zee, Phys.Rev.D19(1979)1036
\bibitem{AD}
I.Affleck and M.Dine, Nucl.Phys.B249(1985)361;
M.Dine, L.Randall, S.Thomas, Nucl.Phys.B458(1996)291.
\bibitem{kusenko}
A.Kusenko, Phys.Lett.B404(1997)285; Phys.Lett.B418(1998)46
\bibitem{Kasuya}
S.Kasuya and M.Kawasaki, Phys.Rev.D61(2000)041301; Phys.Rev.D62(2000)023512.
\bibitem{Enq}
K.Enqvist and J.McDonald, Phys.Lett.B425(1998)309; Nucl.Phys.B538(1999)321.
\bibitem{dineT}
A.Anisimov and M.Dine, ``Some issues in Flat Direction Baryogenesis'',
hep-ph/0008058.
\bibitem{KZO}
Y.B.Zeldovich, I.Y.Kobzarev and L.B.Okun, Zh.Eksp.Teor.Fiz.67(1974)3, 
Sov.Phys.JETP40(1974)1
\bibitem{vilenkin}
A.Vilenkin, Phys.Rev.D23(1981)852
\bibitem{matsuda_tobe}
T.Matsuda, in preparation.
\bibitem{b-l}
M.Fujii, K.Hamaguchi, and T.Yanagida, Phys.Rev.D64(2001)123526.
\bibitem{pati}
J.C.Pati, Phys.Lett.B388(1996)532
\bibitem{z_symmetry}
J.Ellis, A.E.Faraggi, D.V.Nanopoulos, Phys.Lett.B419(1998)123;\\
L.E.Ibanez, F.Quevedo, JHEP 9910(1999)001;\\
A.E.Faraggi, Nucl.Phys.B428(1994)111;\\
J.A.E.Faraggi, Phys.Lett.B388(1996)532.
\bibitem{discrete}
L.E.Ibanez, G.G.Ross, Nucl.Phys.B368(1992)3.
\bibitem{thermal}
D.H.Lyth, E.D.Stewart, Phys.Rev.D53(1996)1784.
\bibitem{matsuda_thermal_wall}
T.Matsuda, Phys.Lett.B486(2000)300.
\bibitem{ellis}
R.Allahverdi, B.A.Campbell, J.R.Ellis, Nucl.Phys.B579(2000)355.
\bibitem{bias}
D.Coulson, Z.Lalak, B.Ovrut, Phys.Rev.D53(1996)4237;\\
Z.Lalak, S.Lola, B.A.Ovrut, G.G.Ross, Nucl.Phys.B434(1995)675
\bibitem{DS}
A.Dolgov and J.Silk, Phys.Rev.D47(1993)4244
\bibitem{simulation}
J.A.Harvey, E.W.Kolb, D.B.Reiss and S.Wolfram,
Nucl.Phys.B201(1982)16;\\
T.Vachaspati and A.Vilenkin, Phys.Rev.D30(1984)2036.
\bibitem{matsuda}
T.Matsuda,Phys.Lett.B436(1998)264.
\bibitem{Abel}
S.A.Abel, S.Sarker and P.L.White, Nucl.Phys.B454(1995)663\\
T.Han, D.Marfatia and Ren-Jie Zhang 
Phys.Rev.D61(2000)013007.
\bibitem{soft_wall}
A.Singh, Phys.Rev.D50(1994)671;\\
A.M.Fuller and D.N.Schramm, Phys.Rev.D45(1992)2595;\\
A.Massarotti and J.M.Quashnock, Phys.Rev.D47(1993)3177;\\
S.Lola and G.G.Ross, Nucl.Phys.B406(1993)452.
\end{thebibliography}
\end{document}